\documentstyle[preprint,aps,tighten]{revtex}

\newcommand\pb{\overline p}
\newcommand\qb{\overline q}
\newcommand\taubar{\overline\tau}

\begin{document}

\thispagestyle{empty}

\title{Transport Properties of the Diluted Lorentz Slab}

\author{
Hern\'an Larralde $^{\star}$, 
Fran\c{c}ois Leyvraz $^{\star}$, \\ 
Gustavo Mart\'{\i}nez-Mekler $^{\star}$, 
Ra\'ul Rechtman $^{\triangle}$\cite{rech}, 
Stefano Ruffo $^{\diamond}$\cite{ruffo}
}

\address{
$\star$ Centro de Ciencias Fisicas, UNAM, Apdo. Postal 48-3, \\
62251 Cuernavaca, Morelos, Mexico \\}

\address{$\triangle$ Centro de Investigaci\'on en Energ\'{\i}a, UNAM \\
62580 Temixco, Morelos, Mexico \\}

\address{$\diamond$ Dipartimento di Energetica ``S. Stecco'', 
Universit\` a di Firenze,
via S. Marta 3, \\
50139 Firenze, Italy, INFM and INFN} 

\maketitle

\begin{abstract}
We study the behavior of a point particle incident from the left on a
slab of a randomly diluted triangular array of circular scatterers.
Various scattering properties, such as the reflection and transmission
probabilities and the scattering time are studied as a function of
thickness and dilution. We show that a diffusion model satisfactorily
describes the mentioned scattering properties.  We also show how some
of these quantities can be evaluated exactly and their agreement
with numerical experiments. Our results exhibit the
dependence of these scattering data on the mean free path. This
dependence again shows excellent agreement with the predictions of a
Brownian motion model.
\end{abstract}

\vspace{0.5cm}
\noindent PACS number(s): 05.40.+j,~05.45.+b,~05.60.+w

\section{Introduction}

The Lorentz gas~\cite{lor05} is one of the fundamental kinetic theory
models~\cite{gallavotti69}, and many of its ergodic properties are well
known~\cite{goldstein75}.  In a previous paper~\cite{larralde98} we
investigated the scattering and transmission properties of an array of
disks centered on a finite triangular lattice, a Lorentz slab.  There,
we studied the validity of approximating the motion of such a particle
by a diffusive process~\cite{bun81}.  Specifically, we considered
samples of finite thickness $L$ along the $x$ axis, but infinite along
the $y$ axis.  We then looked at the behavior of a particle incident on
the sample from the left and measured the probability $T$ of being
transmitted to the right, the reflection probability $R$ and the
average residence time $\langle\tau\rangle$.  We found a very
satisfactory agreement with the predictions of one dimensional Brownian
motion for all these quantities when the distance between scatterers is
small enough to prevent any particle from moving in an  arbitrarily
long straight line (the so-called finite horizon case).  When such
unbounded motion is possible we found that normal diffusive behavior
did not arise.  Rather, a complex pattern of logarithmic corrections
was found for the various quantities of interest. A modification of the
diffusive model considering L\'evy walks as suggested in
Refs.~\cite{bou90,ble92} explains such features.

Here we study the situation in which the system is randomly diluted,
but the scatterers are still placed on the the sites of a triangular
lattice. That is, we consider the case in which a fraction $f$ of the
cells of the periodic array is occupied by scatterers and the rest is
empty.  In such a system, the particle can always take steps of
arbitrary length, but the horizon is said to be finite if the
distribution of free paths has an exponential cutoff.  This happens
when the corresponding system with $f=1$ has finite horizon,{\it i.e.}, the
only large steps that occur in the diluted system  are related to the
absence of a large number of scatterers, which is an exponentially
improbable event. In this paper, we restrict ourselves to this finite
horizon case.

The importance of looking at the randomly diluted case, is that it
provides a controlled way of varying the mean free path---and hence the
diffusion constant---over an arbitrarily large range, without affecting
the property of having a finite horizon and hence a normal diffusive
behavior. This therefore allows us to check the correspondence with
diffusion models much more thoroughly. Further, it shows that the
correspondence between classical deterministic motion and Brownian
motion, as proved rigorously in Ref.~\cite{bun81} for the case of
periodic billiards, does not, in fact, require strict periodicity.

This paper is organized as follows: In Sec.~\ref{models} we describe in
detail how the numerical experiments are performed.  In particular, we
define two different ways in which we introduce disorder in the system,
namely {\it quenched} and {\it annealed}. In Section~\ref{transport},
we discuss the average transport properties, such as transmission
probability, mean free path and average scattering time. We find that
the last two quantities can be evaluated exactly using a relationship
due to Kac~\cite{kac47}.  We display the numerical results for these
quantities as well as the predictions using the Kac formula and the
diffusion model. For the sake of completeness, we reproduce the
derivation of the Kac formula in Appendix A. In Sec.~\ref{distribution}
we discuss the distribution functions of free paths, of residence times
and of heights of exit of transmitted particles. In Appendix B  we
derive the latter two distributions for the diffusion model.  Finally
in Sec.~\ref{conclusions} we present our conclusions.

\section{Model setup}
\label{models}

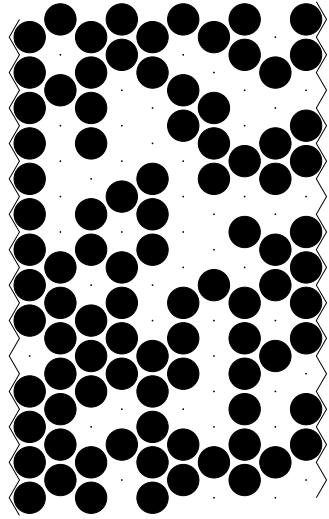
\begin{figure}
\caption{\label{fslab} Slab of diluted scatterers in a triangular array
with $L=10$, $w=0.2$, and $f=0.7$. Point particles enter the slab 
on the left side saw tooth border.} 
\end{figure}

The geometric arrangement of the scatterers in the undiluted ($f=1$) system 
is the following: each scatterer is a disk of unit radius, the centers of 
which form a triangular lattice. The slab is infinite in the vertical 
direction and is characterized by the number $L$ of columns and the minimal 
separation $w$ between the disks. Dilution is then obtained by occupying 
only a fraction $f$ of the sites of the array with disks of unit radius.
A typical scatterer configuration is shown in Fig.~\ref{fslab}.
The left and right ``sawtooth'' borders are the outer sides of the
hexagonal cells attached to each site of the lattice.

Particles are launched from randomly chosen positions along the left
border. Each incident particle has a different impact parameter $b$,
defined here as the distance between the initial position and the
horizontal line passing through the center of the scatterer in the
cell.  The angles of incidence $\theta$ measured with respect to the
side from which the particle is launched, are distributed in the
interval $[0, \pi]$ in such a way as to make $\cos\theta$ uniformly
distributed. This choice reproduces initially the Liouville measure in
the Birkhoff coordinates. The particles move freely except for elastic
collisions at the boundary of the disks.

In the undiluted case ($f=1$), if the separation between scatterers $w$
is small, $0<w<w_c=(4/\sqrt{3}-2)=0.3094\dots$, the length of free
motion of the particles is bounded and they "see" a finite horizon. All
the numerical experiments discussed in this paper are performed in this
range.  On the other hand, in the diluted case, arbitrarily long paths
without collisions can exist also for $0<w<w_c$, but their contribution
to the diffusion constant remains finite.

As far as the dilution process itself is concerned, we realize it in
two different ways corresponding to the usual distinction between {\it
quenched} and {\it annealed} disorder.  The annealed disorder is
obtained by choosing with probability $f$ the cell to be occupied at
the moment at which the particle enters the cell. Thus, when the
particle eventually revisits a given cell, its occupancy status may be
different.  This way of introducing disorder involves a simultaneous
average over the dynamical and the disorder variables.  On the other
hand, we also performed dilution in the more realistic quenched case,
in which a sample is first created, for which all cells are either
occupied or empty with probability $f$ and averages are taken over many
realizations of disorder.

\section{Average transport properties}
\label{transport}

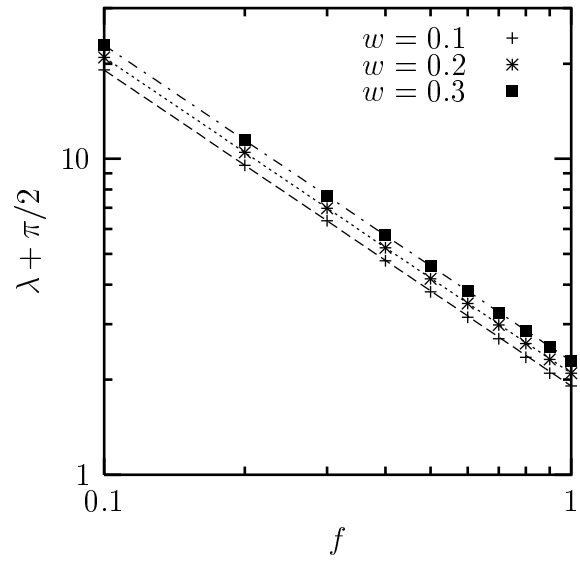
\begin{figure}
\caption{\label{mfp} Mean free path shifted by $\pi/2$ as a function of 
the dilution $f$ for quenched disorder. The curves 
are the r.h.s. of Eq.~(\protect{\ref{eq:mfp}}) again
shifted by $\pi/2$. Each point represents the average over $10^8$ collisions.}
\end{figure}

We begin this Section by showing a derivation of a formula for
the mean free path $\lambda$ as a function of the geometrical
parameter $w$ and dilution $f$. 
This quantity can be evaluated in terms of phase space integrals 
using Kac formula~\cite{kac47}, which assumes ergodicity. Indeed, as 
shown in Appendix A, for a particle moving freely at unit velocity 
in a bounded domain of area $S$, the average return 
time $\langle{\cal T}\rangle$ to a 
boundary segment of length $P$ is  
\begin{equation}
\langle {\cal T} \rangle= \frac{\pi S}{P}.
\label{eq:time}
\end{equation}
For the Lorentz gas, choosing the boundary segment as the perimeter of 
all disks, the average return time coincides with the mean free time 
between two collisions. Since the particles move with unit velocity, 
this also coincides with the mean free path $\lambda$.
Considering a finite array of Wigner-Seitz hexagonal cells with 
$L$ columns and $M$ rows, the total area of the domain is $LMC$, 
where $C=(2+w)^2 \sqrt{3}/2$ is the cell area. From this we must 
subtract the area occupied by the unit radius disks, {\it i.e.} 
$\pi f LM$. The total perimeter is $P=2 \pi f LM$. Hence, from 
Eq.~(\ref{eq:time}) and the above considerations
\begin{equation}
\lambda=\frac{\sqrt3}{4f}(w+2)^2-\frac{\pi}{2}~.
\label{eq:mfp}
\end{equation}
Note that this holds for any value of $M$ and hence
extends trivially to the case we consider, where $M$ is
infinite.  The usual derivation of Eq.~(\ref{eq:time}) 
only applies to the quenched case, {\it i.e.} to a 
scatterer configuration that is fixed in time. However,
the arguments we give in Appendix A show how to extend it to the annealed
case as well. Numerical experiments confirm this finding, for both quenched 
and annealed disorder.
In Fig.~\ref{mfp} we display the results for quenched disorder,
since the annealed data are identical. 

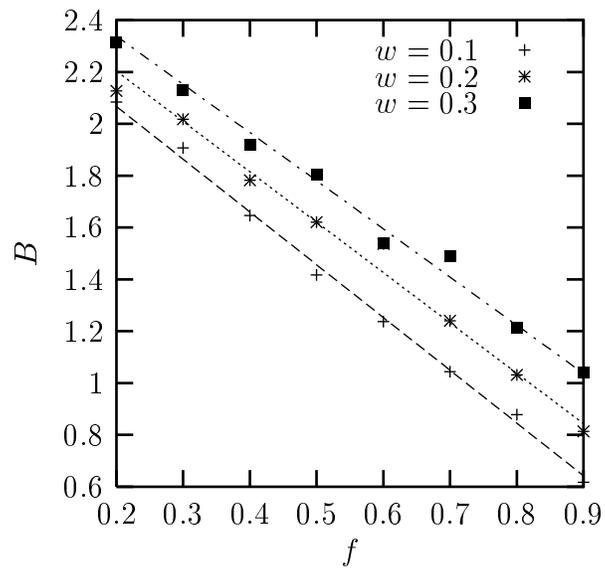
\begin{figure}
\caption{\label{fig:B} Dependence of $B$ on $f$ for quenched disorder. 
The curves are Eq~(\protect{\ref{eq:average-time}}). The values
of $B$ were found by adjusting Eq.~(\protect{\ref{eq:average-time}}) to
the experimental data obtained by letting $10^7$ particles travel
through slabs of lengths going from 100 to 1500.}
\end{figure}

The slab is infinite in the $y$ direction and the collisions are
elastic, hence every particle that enters the slab must eventually leave it,
except for a set of zero measure which goes asymptotically to bounded
orbits inside the slab. Thus, in practice, a particle that enters 
the slab collides with some of the obstacles and is ultimately 
transmitted or reflected. From Eq.~(\ref{eq:time}) one can obtain the 
average residence time $\langle \tau \rangle$ as a function of $w$, $f$ 
and $L$, if one now takes the border of the slab as the boundary segment.
For a slab with $L$ columns and $M$ rows 
the perimeter is $M 4(2+w)/\sqrt{3}$ and the total area of the domain is as 
before. Hence
\begin{equation}
\langle\tau\rangle(f,w,L)=\left( \frac{3 \pi (2+w)}{8} 
- \frac{\sqrt{3} \pi^2 f}{4(2+w)}\right)L=B(f,w) L.
\label{eq:average-time}
\end{equation}
This equation defines the quantity $B$. As above (see Eq.~(\ref{eq:mfp})), 
the result is independent of $M$.
We verified numerically the
linear growth of $\langle\tau\rangle$ with $L$ and found it to be valid,
as expected, also for
small values of $L$. From these experiments we
obtained the values of $B$ shown in Fig.~\ref{fig:B} 
for quenched disorder together with Eq.~(\ref{eq:average-time}).
The agreement in the annealed case (not shown) is even more
satisfactory. Note that $B$ does not diverge as $f \to 0$, 
as opposed to the mean free path. 

Let us now compare these exact results with the predictions of a model 
based on Brownian motion. We assume that the
particles incident from the left penetrate a given
distance $a$ into the sample, after which they forget
everything about the way they were injected into the system
and diffuse with diffusion constant $D$.
Under these assumptions, the quantities of interest
can be evaluated exactly in terms of $a$ and $D$ and 
compared with the exact results obtained above. Thus, the
average time to reach either side starting at a distance
$a$ from the left side is given by~\cite{fel,weiss}
\begin{equation}
\langle\tau\rangle(L)=\frac{a(\ell L-a)}{2D}~,
\label{eq:av-time-diff}
\end{equation}
where $\ell=\sqrt{3}(1+w/2)$ is the horizontal separation 
between columns
of the slab. One sees therefore that the form of the $L$ dependence 
in the diffusive case is slightly different from the exact one, 
Eq.~(\ref{eq:average-time}), since in that case the proportionality 
to $L$ is exact over the whole range of $L$. On the 
other hand, in Eq.~(\ref{eq:av-time-diff}), a constant 
term appears which is negligible in the limit 
$a/\ell\ll L$. Since this is the limit for which diffusion
is expected to be a valid description, this is not a serious
problem. On  the 
other hand, this allows to evaluate the ratio $a/D$ exactly 
in terms of $f$ and $w$, via
\begin{equation}
\frac{a}{D}=\frac{2B(f,w)}{\ell}.
\label{eq:ratio-a-D}
\end{equation}
It is well known that it is not possible to obtain exact evaluations
of the same sort for the diffusion constant $D$, since
its value can be shown to depend in a detailed manner on
the specific dynamics involved. Indeed, 
\begin{equation}
D=\lim_{\epsilon\to0}\int_0^\infty 
\overline{\dot x(0)\dot x(t)}e^{-\epsilon t}dt,
\end{equation}
where the bar denotes an average over the realizations
of the disorder.
From this it follows that an analytical expression for $D$
in terms of simple phase space expressions is impossible.
However, an order of magnitude estimate for $D$ can be
given. Since the mean square distance
grows linearly with time, and since the only microscopic 
length scale is the mean free path, we are led to estimate
$Dt_0$ by the square of the mean free path, where $t_0$ is
the time needed for the particle to cover a mean free path.
In our system, the velocity of the particle is
constant and equal to one, therefore the result is that $D$
is of the same order as $a$, both being of the order of the
mean free path. This is indeed consistent with 
Eq.~(\ref{eq:ratio-a-D}).  It should be noted that the ratio 
of the two dynamics dependent quantities $a$ and $D$ depends 
only on the geometric features of the system.

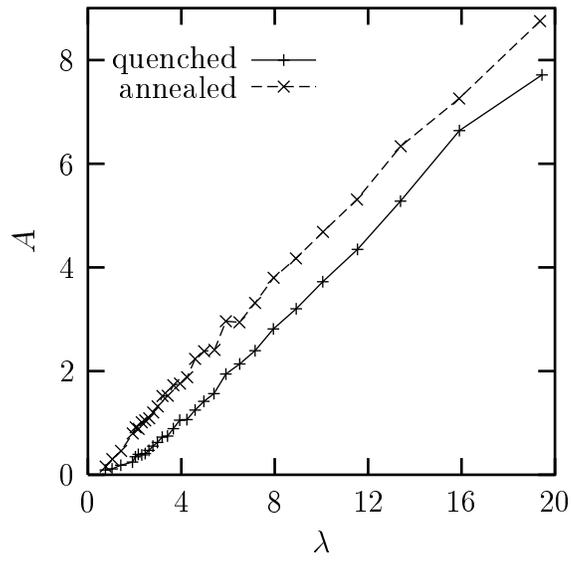
\begin{figure}
\caption{\label{fig:A} Dependence of $A$ 
on the mean free path $\lambda$
for $w=0.2$. Both quenched and annealed disorder are shown. The values
of $A$ were found by adjusting Eq.~(\protect{\ref{eq:def-A}}) to
the experimental data obtained by letting $10^7$ particles travel
through slabs of lengths going from 100 to 1500.}
\end{figure}

Let us now turn to another average transport property, 
for which no exact expression is available, namely the
transmission probability $T$ as a function of $L$. One
finds that asymptotically, in the limit $L \gg 1$, 
\begin{equation}
T(f,w,L)=\frac{A(f,w)}{L}
\label{eq:def-A}
\end{equation}
In Fig.~\ref{fig:A}, we show that $A$ grows almost linearly with
$\lambda$, for large values of $\lambda$. This can readily be
understood in terms of the diffusive model. Indeed, in the case of a
diffusing particle being injected at a distance $a$ from the left-hand
side of the slab, the probability that it be transmitted to the
right-hand side without being absorbed first at the left-hand side
$T_d$  is well-known to be~\cite{fel,weiss}
\begin{equation}
T_d(L)=\frac{a}{\ell L}.
\end{equation}
From this we see that $A$ can be identified with a quantity which, as
argued above, is expected to scale as the mean free path. Note further
that whereas $\langle\tau\rangle$ and $\lambda$ are identical in the
quenched and annealed cases, this is not true for the transmission
probability, which shows significant differences for the two forms of
disorder.

\section{Distribution functions}
\label{distribution}

We have also studied the distributions of residence times of 
transmitted (respectively reflected) particles, of free paths and
of the heights at which transmitted (respectively
reflected) particles leave the system.

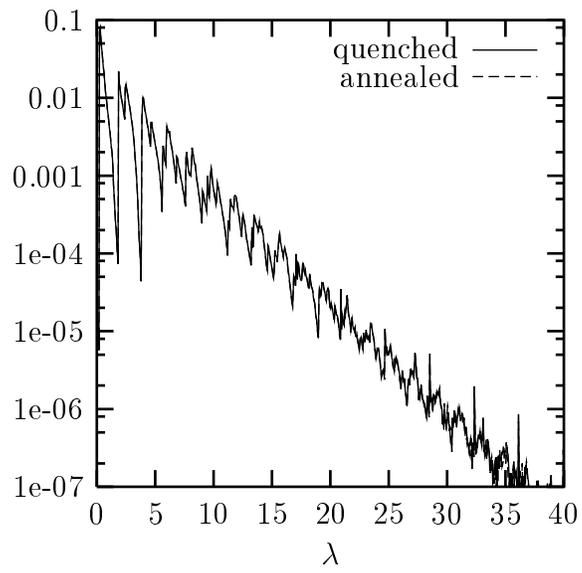
\begin{figure}
\caption{\label{mfp-dist} Free path distribution for $w=0.2$,
$f=0.5$ for both quenched and annealed disorder. The curves are
practically superimposed. The distributions were found from $10^8$ collisions
in each case.}
\end{figure}

Of these, only the distribution of free paths is accessible to exact
theoretical treatment. Indeed, this is nothing but an equilibrium
property, and can in principle be evaluated using an integral over the
perimeter of the billiard with the Birkhoff measure.  However, this is
not really practicable, so we have not attempted it.  The distribution
of mean free paths is quite irregular (see Fig.~\ref{mfp-dist}). This
irregularity arises from the presence of resonances and the fact that
there are forbidden distances.  Note that, as remarked in the
Introduction, the distribution decays exponentially since the horizon
is finite.  From the above characterization of the mean free path
distribution as an equilibrium property, it readily follows that they
are independent of whether the average is taken as quenched or
annealed.

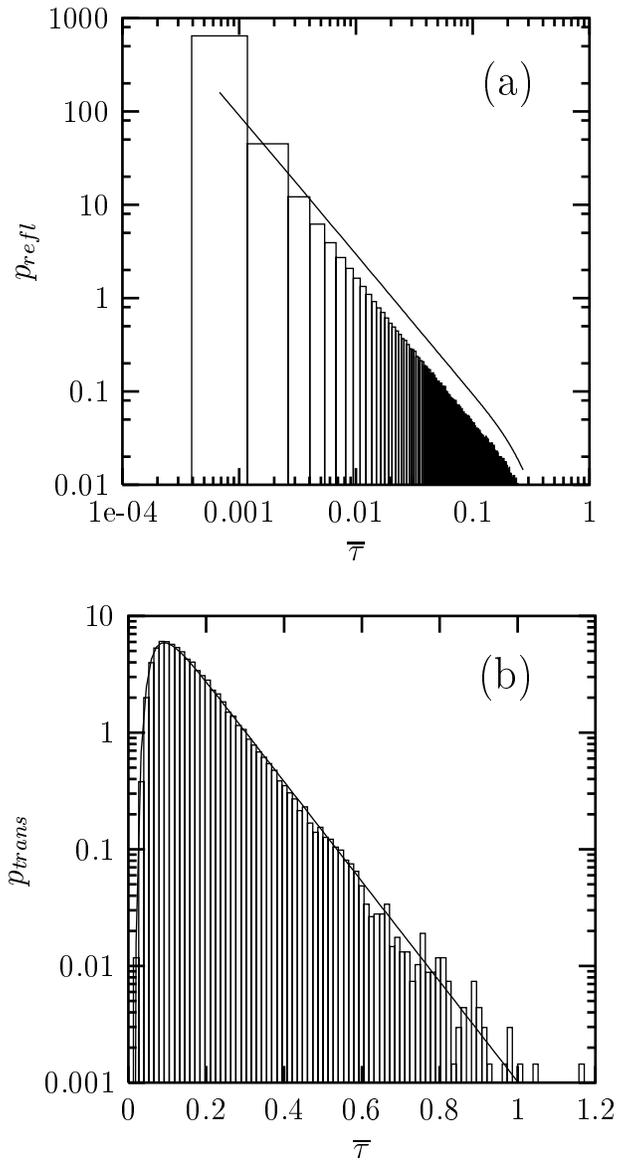
\begin{figure}
\caption{\label{fig:time-distribution} 
Distribution of residence times for 
reflected (a) and transmitted (b) particles for
quenched disorder, $w=0.2$ and $f=0.5$. The continuous curve 
represents the fit described in the text. The experimental 
distributions were found by letting $10^7$ particles travel 
through a slab with 100 columns.}
\end{figure}

Now let us consider the distribution of residence times,  which is not
susceptible of an exact evaluation.  However, it can be computed in the
diffusion model. It is clearly sufficient to compute the distribution
of residence times for transmitted particles $p_{trans}$ upon starting
at $a$, since the corresponding distribution for reflected particles
$p_{refl}$ arises by substituting $a$ by $\ell L-a$ in $p_{trans}$.
Defining the scaled variables $\taubar=D\tau/(\ell L)^2$ and
$\alpha=a/(\ell L)$, the final result is, as shown in Appendix
B~\cite{weiss}
\begin{equation}
p_{trans}(\taubar;\alpha)=2\pi\sum_{n=1}^\infty%
n\sin \pi n(1-\alpha)\,e^{-\pi^2 n^2\taubar}.
\label{eq:time-distribution}
\end{equation}
Integrating, one obtains Eq.~(\ref{eq:av-time-diff}) 
for the average time
a diffusing particle takes to reach either side. 
We fit this to the numerical data as follows: one first 
considers the average time that a transmitted particle remains
in the sample. This is obtainable from the experimental distribution
on the one hand, but is also given by
\begin{equation}
\langle\tau\rangle_{trans}=\frac{\ell^2L^2}{6D}.
\end{equation}
This allows therefore a determination of $D$ from the data, which for $f=1$
is in good agreement with published data~\cite{del00}.
From this and Eq.~(\ref{eq:average-time}) together 
with Eq.~(\ref{eq:av-time-diff}), one also obtains a value of the
parameter $a$. One is then in a position to plot the
theoretical curve Eq.~(\ref{eq:time-distribution}) together
with the empirical data. This is shown in 
Fig.~\ref{fig:time-distribution}, for an occupation fraction 
$f=0.5$. The agreement is quite good. The apparent 
shift between the theoretical curve and the data in reflection
can be traced back to an issue of normalization
involving particles reflected after
a very few bounces, which therefore do not show
diffusive behavior. Thus, the empirical distribution 
has a short-time cut-off at larger times than the corresponding 
diffusive model. 

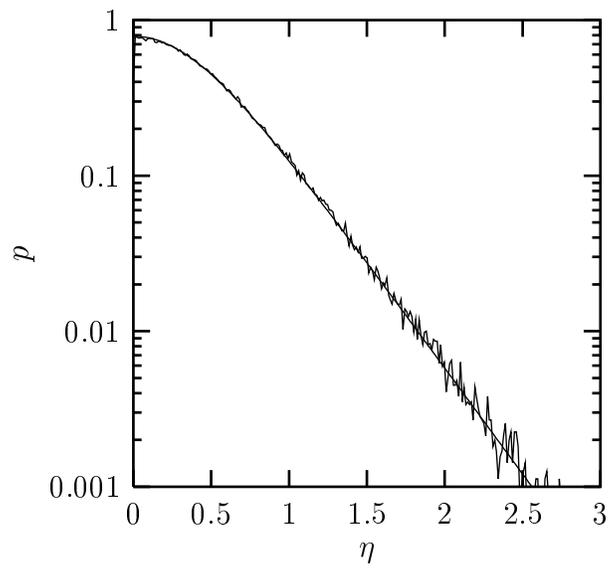
\begin{figure}
\caption{\label{height-dist} 
Height distribution of transmitted particles for quenched
disorder, $w=0.2$ and $f=0.2$. The continuous curve is the plot
of Eq.~(\protect{\ref{eq:height-distribution}}). The experimental 
distribution was found by letting $10^7$ particles travel through a slab 
with 100 columns.}
\end{figure}

Finally, we measure the distribution of heights of 
transmitted particles. This distribution can be
computed in the diffusive model. The result 
for transmitted particles is, as shown in Appendix B,
\begin{equation}
p(\eta)=\frac{\sin\pi(1-\alpha)}{2\alpha}
\frac{1}{\cosh\pi\eta-\cos\pi
(1-\alpha)},
\label{eq:height-distribution}
\end{equation}
where $\eta$ is the scaled height $y/(\ell L)$ and $\alpha$ is as above. 
In Fig.~\ref{height-dist} we show $p(\eta)$ for transmitted particles.
The agreement is excellent. We do not show the behavior for reflected particles
as it is dominated by rapid reflections.

\section{Conclusions}
\label{conclusions}
Summarizing, we have studied transport properties of finite size
samples of Lorentz gases in a situation in which the mean free path can
be varied over a large range of values without affecting the property
of normal diffusion. The mean free path was varied over a considerable
range (slightly more than an order of magnitude), and the agreement
with the Brownian motion model was satisfactory throughout. We have
also shown how the diffusion constant and the penetration depth, which
were the two dynamical parameters of our model, vary with dilution, and
hence with the mean free path.

\section*{Acknowledgments}
We thank A. Bufetov for stimulating discussions.
FL acknowledges the hospitality of the 
Dipartimento di Energetica of the University of Florence, Italy,
where part of this work was carried out.
This work has been partially supported by INFN, INFM, University of 
Florence, MURST-COFIN00 grant on {\it Chaos and Localization in 
Classical and Quantum Mechanics\/}, DGAPA-UNAM under grants IN103595, 
IN106597, IN112200 and CONACyT under grants 32418-E and 32173-E.

\appendix
\section{}
In this Appendix, we show in detail how the 
exact relations Eq.~(\ref{eq:mfp}) and 
Eq.~(\ref{eq:average-time}) are obtained. 
To this end, we first recall the derivation of a general 
formula due to Kac. 
Consider a $(2N-1)$-dimensional energy shell in a $2N$-dimensional
Hamiltonian phase-space and select a $(2N-2)$-dimensional 
Poincar\'e surface that intersects all (or nearly all) 
trajectories on the energy shell.
If such a surface cannot be found, then the phase space integrals
below must be restricted to that part of the phase space which can 
be reached from the surface.

Parametrize each point $(p,q)$ on the energy surface using the
last point $(\pb,\qb)$ on the Poincar\'e surface that lies 
on the trajectory
passing through $(p,q)$. Denote by ${\cal T}_E(\pb,\qb)$
the time necessary to reach $(p,q)$ starting from $(\pb,\qb)$.
This defines the canonical coordinate transformation
\begin{equation}
d^Np\,d^Nq=d^{N-1}\pb\, d^{N-1}\qb\,d{\cal T}\,dE.
\end{equation}
The constant energy $E_0$ volume is hence
\begin{equation}
\int d^N p\, d^Nq\,\delta[E_0-H(p,q)]=
\int d^{N-1}\pb \, d^{N-1} \qb\,{\cal T}_{E_0}(\pb,\qb).
\end{equation}
Denoting by ${\cal N}$, the total $(2N-2)$-dimensional 
phase space volume of the Poincar\'e surface, 
one immediately obtains the Kac formula
for the average time to return to the surface:
\begin{eqnarray}
\label{eq:average}
\langle{\cal T}_{E_0}\rangle&=&{\cal N}^{-1}\int\delta
[E_0-H(p,q)]d^Np\,
d^Nq,\\
{\cal N}&=&\int_{E=E_0}d^{N-1} \pb d^{N-1} \qb. \nonumber
\end{eqnarray}
To evaluate these integrals in the case of billiards for 
$N=2$ (of which the Lorentz gas 
with quenched disorder is a particular case), we take the
Hamiltonian to be $p^2/2$ and $E_0$ to be $1/2$. From this follows
\begin{equation}
\int d^2 p\,d^2q\, \delta[1/2-p^2/2]=2\pi S,
\end{equation}
where $S$ is the area of the billiard.
As a Poincar\'e surface we introduce an arbitrary 
subset of the billiard boundary having perimeter $P$, 
with the usual Birkhoff coordinates as variables 
$\pb$ and $\qb$. One then finds
\begin{equation}
\int_{E=E_0}d \pb\,d \qb= P \int_{-\pi/2}^{\pi/2}\cos\theta\,
d\theta= 2P,
\end{equation}
From these two equations one derives Eq.~(\ref{eq:time})

The above remarks are clearly limited to the case of quenched disorder.
The generalization to the annealed case can be made as follows: to
every point $(\pb,\qb)$ on the Poincar\'e surface, add a doubly
infinite sequence of zeroes and ones
$\left(\sigma_k\right)_{k=-\infty}^\infty$, which we denote by
$\overline\sigma$, with all $\sigma_k$ independently distributed and
equal to one with probability $f$.  We now define the dynamics as
follows: The orbit starts from $(\pb,\qb)$ and the cell at $t=0$ is
occupied or empty according to the value of $\sigma_0$. The orbit then
proceeds until it leaves the cell. The status of the next cell is then
decided according to the value of $\sigma_1$ and so on. Although
discontinuities arise when a trajectory crosses a vertex of the
fundamental cell, the dynamics is still given by a canonical map. Thus,
for a {\it fixed\/} sequence $\overline\sigma$ the formula
Eq.~(\ref{eq:average}) applies. However, for a fixed sequence, we
cannot easily compute the phase space volumes involved.  Since we are
only interested in the average of $\langle{\cal
T}\rangle(\overline\sigma)$ over all values of $\overline\sigma$, we
note the following: the denominator in Eq.~(\ref{eq:average}) is
independent of $\overline\sigma$, so that it is sufficient to average
over the numerator.  This average can be performed separately for each
cell, and the result therefore follows trivially.

\section{}
In this Appendix, we derive Eq.~(\ref{eq:time-distribution})
and Eq.~(\ref{eq:height-distribution}).
To this end, we first derive an analogue of
Eq.~(\ref{eq:time-distribution})
for the probability  that a particle first exits the slab on the right 
side at a height $y$ and at time $t$. In order to simplify the notation, we
first go over to scaled variables $\xi=x/(\ell L)$, 
$\eta=y/(\ell L)$,
$\alpha=a/(\ell L)$ as well as $\taubar=Dt/(\ell L)^2$. 
All distribution functions are further rescaled
in such a way as to remain normalized.
In these variables, this probability is given
by the following expression:
\begin{equation}
p_{trans}(\eta;\taubar)=-\left.\frac{\partial P_0(\xi.\eta;\taubar)}
{\partial\xi}\right|_{\xi=1},
\end{equation}
where $P_0(\xi,\eta)$ is the solution of the following problem:
\begin{eqnarray}
\frac{\partial P_0(\xi,\eta;\taubar)}{\partial \taubar}
&=&\Delta P_0(\xi,\eta),\nonumber\\
P_0(\xi,\eta;0)&=&\delta(\xi-\alpha)\delta(\eta)
\qquad P_0(0,\eta;\taubar)=P_0(1,\eta;\taubar)=0.
\label{eq:problem}
\end{eqnarray}
This is now solved by developing the delta function in eigenmodes
of the Laplacian satisfying the boundary
conditions in Eq.~(\ref{eq:problem}), that is
\begin{equation}
P_0(\xi, \eta;\taubar)=2\sum_{n=1}^\infty\sin n\pi\xi
\sin n\pi\alpha
\int_{-\infty}^\infty\frac{dk}{2\pi}e^{ik\eta}
\exp\left[-(\pi^2n^2+k^2)\taubar\right].
\end{equation}
The  resulting expression for $p_{trans}(\eta,\taubar)$
can now be integrated either over $\eta$ or over $\taubar$
to yield Eq.~(\ref{eq:time-distribution}) or
Eq.~(\ref{eq:height-distribution}) respectively. 
In either case, obtaining the results in the text is 
now a matter of straightforward algebra.

\end{document}